\documentclass{basi}
\usepackage[british]{babel}
\usepackage[varg]{txfonts}

\begin{document}

\title[Modelling solar cycle]{Double ring algorithm of solar active region eruptions within the framework of kinematic dynamo model}
\author[s hazra]{Soumitra Hazra$^1$\thanks{email: \texttt{s.hazra@iiserkol.ac.in}} \&
Dibyendu Nandy$^{1}$\thanks{email: \texttt{dnandi@iiserkol.ac.in}} \\
$^1$Department of Physical Sciences, Indian Institute of Science Education and Research, Kolkata, Mohanpur, 741252, India}

\date{Received --- ; accepted ---}

\maketitle
\label{firstpage}
\begin{abstract}
Recent results indicate that the Babcock-Leighton mechanism for
poloidal field creation plays an important role in the solar cycle.
However, modelling this mechanism has not always correctly captured the
underlying physics. In particular, it has been demonstrated that using a
spatially distributed near-surface alpha-effect to parametrize the
Babcock-Leighton mechanism generates results which do not agree with
observations. Motivated by this, we are developing a physically more
consistent model of the solar cycle in which we model poloidal field
creation by the emergence and flux dispersal of double-rings structures.
Here we present preliminary results from this new dynamo model.
\end{abstract}
\section{Introduction}

It is observed that the number of sunspots on the solar surface varies periodically on the solar surface with a periodicity of 11 years. One of the most promising approaches for modelling this solar cycle is using axisymmetric kinematic solar dynamo models. In such models, toroidal field generation takes place in the tachocline by stretching of poloidal field lines in $\phi$ direction by the differential rotation, while poloidal field generation takes place due to decay of tilted bipolar sunspot pairs near solar surface, commonly known as Babcock-Leighton mechanism. The widely used approach of treating the Babcock-Leighton mechanism by a spatially distributed near-surface alpha effect is computationally convenient but has some inherent problems. Here we develop a model for the solar cycle, where we model the generation of poloidal fields by the emergence and flux dispersal of double ring structures.

\section{Model and Results}

Here we assume axisymmetry in our calculations. The axisymmetric dynamo equations are\\
\begin{equation}
   \frac{\partial A}{\partial t} + \frac{1}{s}\left[ \textbf{v}_p \cdot \nabla (sA) \right] = \eta\left( \nabla^2 - \frac{1}{s^2}  \right)A + K_0f(r,\theta)F(B_{tc})B_{tc}
\end{equation}\\
\begin{equation}
   \frac{\partial B}{\partial t}  + s\left[ \textbf{v}_p \cdot \nabla\left(\frac{B}{s} \right) \right] + (\nabla \cdot \textbf{v}_p)B = \eta\left( \nabla^2 - \frac{1}{s^2}  \right)B + s\left(\left[ \nabla \times (A\bf \hat{e}_\phi) \right]\cdot \nabla \Omega\right)   + \frac{1}{s}\frac{\partial (sB)}{\partial r}\frac{\partial \eta}{\partial r}
\end{equation}\\
where A is the poloidal field, B is the toroidal field , $v_p$ is the meridional flow, $\Omega$ is the differential rotation, $\eta$ is the turbulent magnetic diffusivity and $s = r\sin \theta$. For simulations with double ring algorithm, $K_0 =0$. \\
 In this model we use a two step radially dependent magnetic diffusivity profile as described in Mu{\~n}oz-Jaramillo, Nandy \& Martens (2009). In our case, the diffusivity at the bottom of convection zone is $10^8 cm^2/s$, diffusivity in the convection zone is $10^{11} cm^2/s$ and supergranular diffusivity is $5 \times 10^{12} cm^2/s$. $r_{cz} = 0.73R_\odot$, $d_{cz} = 0.025R_\odot$, $r_{sg} = 0.95R_\odot$ and $d_{sg} = 0.015R_\odot$ describes the transition from one diffusivity value to another. We also use same differential rotation profile as described in Mu{\~n}oz-Jaramillo, Nandy \& Martens (2009).\\
 We generate the meridional circulation profile ($v_p$) for a compressible flow inside the convection zone from this equation
  $\nabla.(\rho v_p)=0$      so,       $\rho v_p = \nabla \times (\psi \hat{e_\phi})$\\
   where $\psi$ is\\
  \begin{equation}
   \psi r \sin \theta = \psi_0 (r - R_p) \sin \left[ \frac{\pi (r - R_p)}{(R_s - R_p)} \right] \{ 1 - e^{- \beta_1 r \theta^{\epsilon}}
\}\{1 - e^{\beta_2 r (\theta - \pi/2)} \} e^{-((r -r_0)/\Gamma)^2}
\end{equation}
 Here $\psi_0$ is the factor which determines the the maximum speed of the flow. We use the following parameter values $ \beta_1=1.5, \beta_2=1.8, \epsilon=2.0000001, r_0=(R_0-R_b)/4, \Gamma=3.47 \times 10^8 , \gamma=0.95, m=3/2$. Here $R_p=0.65R_0$ is the penetration depth of the meridional flow. In this present work we use the surface value of meridional circulation as 14m/sec.
 
 \subsection{Modelling Active Regions as Double Rings}
  It is believed that poloidal field generation inside the sun takes place through the Babcock-Leighton effect and helical turbulence. Here we only consider the Babcock-Leighton alpha effect, wherein, poloidal field generation takes place due to decay of bipolar sunspot region. To model active region emergence, we follow the methods of double ring proposed by Durney (1997), Nandy and Choudhuri (2001), Mu\~noz-Jaramillo, Nandy, Martens \& Yeates (2010) and Nandy, Mu\~noz-Jaramillo \& Martens (2011). We define the $\phi$ component of potential vector A corresponding to active region as
  \begin{equation}\label{Eq_AR}
    A_{ar}(r,\theta)= K_1 A(\Phi)F(r)G(\theta),
\end{equation}
where $K_1$ is a constant which ensures super-critical solutions and strength of ring doublet is defined by  $A(\Phi)$ . We define $F(r)$  as
\begin{equation}
    F(r)= \left\{\begin{array}{cc}
            0 & r<R_\odot-R_{ar}\\
            \frac{1}{r}\sin^2\left[\frac{\pi}{2 R_{ar}}(r - (R_\odot-R_{ar}))\right] & r\geq R_\odot-R_{ar}
          \end{array}\right.,
\end{equation}
where $R_\odot$ is the solar radius and $R_{ar}=0.85R_\odot$ corresponds to the penetration depth of the AR. We define   $G(\theta)$  in integral form as
\begin{equation}
    G(\theta) = \frac{1}{\sin{\theta}}\int_0^{\theta}[B_{-}(\theta')+B_{+}(\theta')]\sin(\theta')d\theta',
\end{equation}
where $B_{+}$ ($B_{-}$) defines the strength of positive (negative) ring:
 \begin{equation}\label{Eq_AR_Dp}
    B_{\pm}(\theta)= \left\{\begin{array}{cc}
                     0 & \theta<\theta_{ar}\mp\frac{\chi}{2}-\frac{\Lambda}{2}\\
                     \pm\frac{1}{\sin(\theta)}\left[1+\cos\left(\frac{2\pi}{\Lambda}(\theta-\theta_{ar}\pm\frac{\chi}{2})\right)\right] & \theta_{ar}\mp\frac{\chi}{2}-\frac{\Lambda}{2} \leq \theta < \theta_{ar}\mp\frac{\chi}{2}+\frac{\Lambda}{2}\\
                     0 & \theta \geq \theta_{ar}\mp\frac{\chi}{2}+\frac{\Lambda}{2}
               \end{array}\right..
\end{equation}
Here $\theta_{ar}$ is emergence co-latitude, $\Lambda$ is the diameter of each polarity of the double ring  and $\chi = \arcsin[\sin(\gamma)\sin(\Delta_{ar})]$ is the latitudinal distance between the centres, where the angular distance between polarity centres $\Delta_{ar}=6^o$ and the AR tilt angle is $\gamma$.  We set $\Lambda$, i.e diameter of ring doublet as $6^o$.
\begin{figure}
\begin{center}
\begin{tabular}{cc}
\includegraphics[scale=0.25]{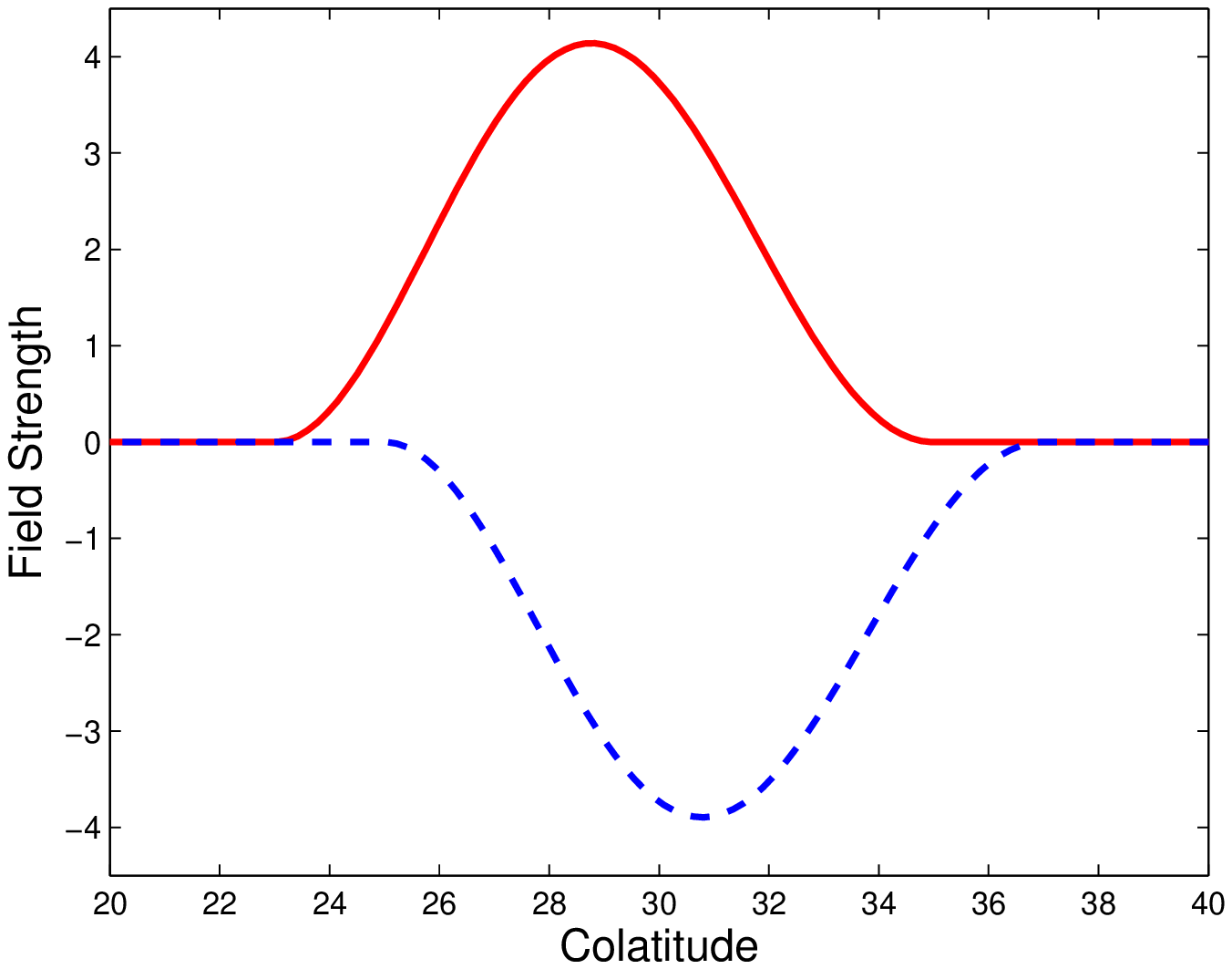} & \includegraphics[scale=0.25]{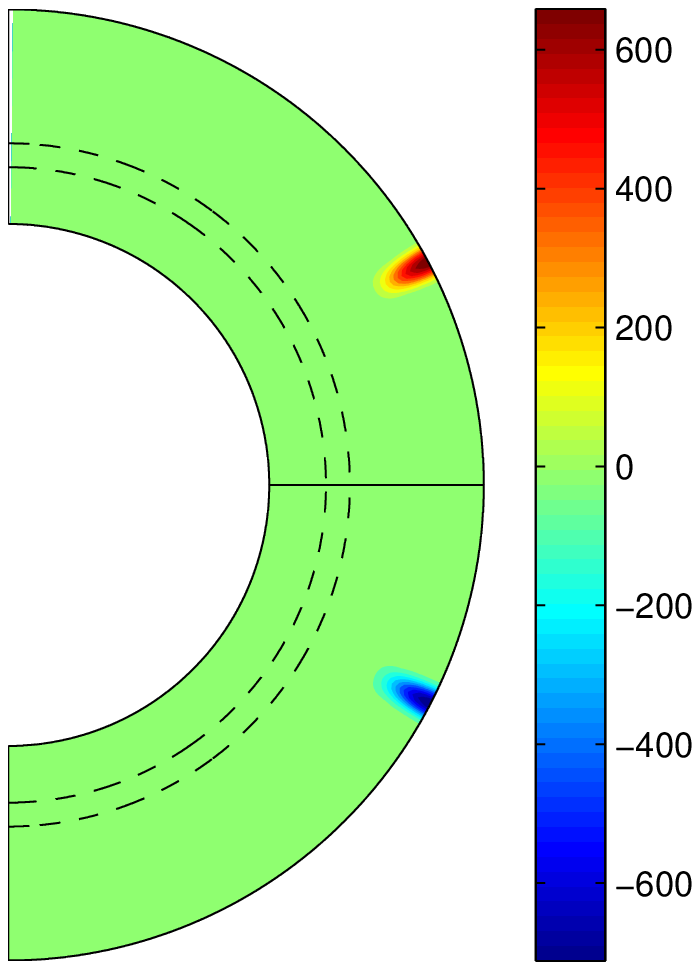}\\
                (a)                                           &                   (b)\\
 \end{tabular}
\end{center}               
\caption{ \footnotesize (a) Diagram illustrating the quantities which define the latitudinal dependence of a double-ring bipolar pair.  (b) Poloidal field lines of our double-rings, one is in northern hemisphere and other is in southern hemisphere.}
\end{figure}

\subsection{Recreating the Poloidal Field}\label{sec_dblr2pld}

In this model, first we check where the toroidal field is higher than the buoyancy threshold at the bottom of convection zone in both northern and southern hemispheres. Then we choose one of the latitude randomly from both the hemisphere simultaneously at a certain interval of time, using a non uniform probability distribution function such that randomly chosen latitudes remain within the observed active latitudes. The probability distribution function is made to drop steadily to zero between 30$^o$ (-30$^o$) and 40$^o$ (-40$^o$) in the northern (southern) hemisphere. Second we calculate the magnetic flux of this toroidal ring.
Then we find tilt of corresponding active region, using the expression given in Fan, Fisher \& McClymont (1994\nocite{fan-fisher-mcclymont94})
\begin{equation}\label{Eq_Prob}
   \gamma \propto \Phi_0^{1/4}B_0^{-5/4}\sin(\lambda),
\end{equation}
where $B_0$ is the local field strength, $\Phi_0$ is the flux associated with the toroidal ring and $\lambda$ is the emergence latitude. We set the constant such that tilt angle lies between  3$^o$ and  12$^o$. \\ Third, we remove a chunk of magnetic field with same angular size as the emerging AR from this toroidal ring and calculate the magnetic energy of the new partial toroidal ring. Then we fix the value of toroidal field such that the energy of the full toroidal ring filled with new magnetic field strength is the same as the magnetic field strength for the partial toroidal ring. Finally, we place the ring duplets with these calculated properties at the near-surface layer at the latitudes where they erupt.
 
 When we solve these equations with appropriate boundary condition, we can successfully simulate the 11 year solar cycle and reproduce the properties of the observed solar butterfly diagram.
 
\begin{figure}[!h]
 \begin{tabular}{c}
\includegraphics[width=13.0cm,height=4cm]{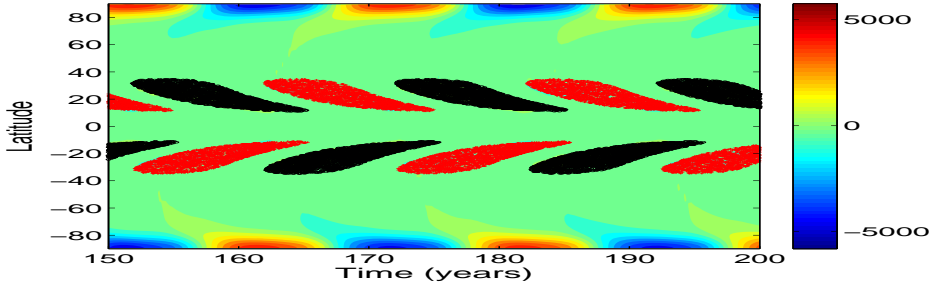}
\end{tabular}
\caption{ \footnotesize  The background shows weak diffuse radial field on solar surface. Eruption latitudes are denoted by symbols black (“o”) and red (“+”), indicating underlying negative and positive toroidal field respectively.}
 \end{figure}

We anticipate that solar dynamo models based on this double ring algorithm would be quite useful in simulating and understanding various features of the solar cycle. \\
Acknowledgements: We thank the University Grants Commission, the Council for Scientific and Industrial Research and the Ramanujan fellowship of the Department of Science and Technology of the Government of India for supporting our research.

\end{document}